\documentclass[aps,twocolumn,pre]{revtex4-2}

\usepackage{xcolor,hyperref}
\hypersetup{
   colorlinks,
   linkcolor={blue!50!black},
   citecolor={blue!50!black},
   urlcolor={blue!80!black}
}

\usepackage{epsfig, amsmath}
\usepackage{bm}
\usepackage{soul}
\usepackage{hyperref}
\usepackage{footmisc}
\usepackage{wrapfig}
\DeclareMathAlphabet{\mathitbf}{OML}{cmm}{b}{it}

\renewcommand{\=}{\!=\!}
\newcommand{\dbar}{{\,\mathchar'26\mkern-12mu d}}

\begin{document}

\title{Nonphononic spectrum of two-dimensional structural glasses}
\author{Edan Lerner$^{1}$}
\email{e.lerner@uva.nl}
\author{Eran Bouchbinder$^{2}$}
\email{eran.bouchbinder@weizmann.ac.il}
\affiliation{$^{1}$Institute of Theoretical Physics, University of Amsterdam, Science Park 904, 1098 XH Amsterdam, the Netherlands\\
$^{2}$Chemical and Biological Physics Department, Weizmann Institute of Science, Rehovot 7610001, Israel}

\maketitle

Extensive numerical work from recent years has established that the low-frequency nonphononic vibrational spectrum of structural glasses follows a $\sim\!\omega^4$ scaling with angular frequency $\omega$~\cite{JCP_Perspective}. This universal quartic ($\sim\!\omega^4$) law featured by the nonphononic spectrum has been shown to be independent of details of the interparticle interaction potential~\cite{modes_prl_2020}, glass formation protocol~\cite{pinching_pnas}, and spatial dimensions $\dbar\!\ge\!2$~\cite{modes_prl_2018,Atsushi_high_D_packings_pre_2020}. Furthermore, indirect evidence for the quartic scaling of the nonphononic spectrum of two-dimensional structural glasses has been presented in~\cite{lte_pnas,cge_paper,JCP_Perspective,jcp_letter_scattering_2021}. In~\cite{modes_prl_2018,cge_paper}, it has been shown that in two-dimensional structural glasses, the prefactor $\omega_{\rm g}^{-5}$ of the quartic $\sim\!\omega^4$ law is $N$-dependent, scaling as $\omega_{\rm g}^{-5}\!\sim\!(\sqrt{\log N})^5$, where $N$ denotes the number of particles in a glass. 

At the same time, deviations from the quartic  scaling of the nonphononic spectrum have been reported for three-dimensional (3D) structural glasses in~\cite{protocol_prerc,lerner2019finite}, depending on the system size and formation protocol of the glasses studied. These previous works associated these deviations with a glassy length $\xi_{\rm g}$ --- on the order of a few interparticle distances ---, and established the nonphononic spectrum of glasses whose linear size sufficiently exceeds $\xi_{\rm g}$ features the universal quartic law $\sim\!\omega^4$~\cite{modes_prl_2018,JCP_Perspective}. The length $\xi_{\rm g}$ has been shown to be glass-formation-protocol dependent~\cite{pinching_pnas} --- decreasing for lower-energy, more stable glasses --- fully consistent with previously observed deviations~\cite{lerner2019finite,protocol_prerc} in the exponent $\beta$ of the nonphononic spectrum $\sim\!\omega^\beta$ from the apparently universal value $\beta\!=\!4$~\cite{JCP_Perspective}.

Recently, in~\cite{grzegorz_2d_modes_prl_2021,grzegorz_erratum_2022} it was argued that the nonphononic spectrum of \emph{two-dimensional} (2D) structural glasses rather follows $\sim\!\omega^\beta$ with $\beta\!<\!4$, presumably casting doubt on the validity of previous observations and claims~\cite{modes_prl_2018,lte_pnas,cge_paper,JCP_Perspective}. Here, we study the nonphononic spectrum of two-dimensional structural glasses, and show that ($i$) the exponent $\beta$ is glass-formation-protocol \emph{and} system-size dependent, as seen previously for structural glasses in three dimensions~\cite{protocol_prerc, lerner2019finite}, and ($ii$) the scaling $\omega_{\rm g}\!\sim\!1/\sqrt{\log N}$ put forward in~\cite{cge_paper,modes_prl_2018} is consistent with numerical observations presented below. 

\begin{figure*}[ht!]
  \includegraphics[width = 0.75\textwidth]{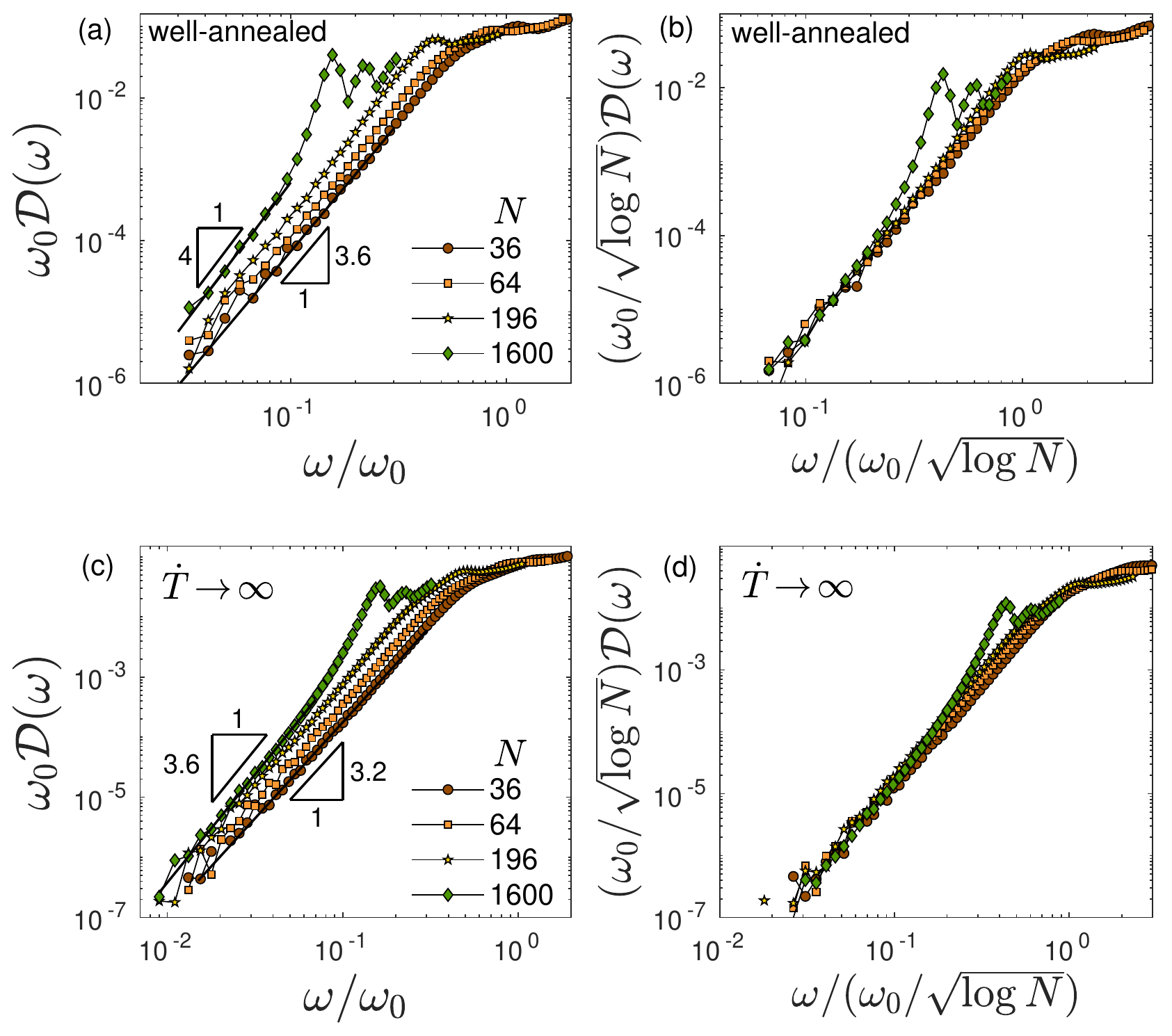}
  \caption{\footnotesize (a) The VDoS ${\cal D}(\omega; N)$ of well-annealed computer glasses (see text for details) vs.~$\omega/\omega_0$, for various system sizes $N$ (see legend), where $\omega_0\!\equiv\!c_s/a_0$ with $c_s$ being the shear wave speed and $a_0$ an interparticle distance. Note that the 4:1 and 3.6:1 scaling triangles correspond to the lines going through the $N\!=\!1600$ and $N\!=\!36$ data, respectively, demonstrating a systematic variation of the exponent $\beta$ with $N$. (b) The same data as in panel (a), but here plotting $(\omega_0/\!\sqrt{\log{N}})\,{\cal D}(\omega; N)$ vs.~$\omega/(\omega_0/\!\sqrt{\log{N}})$. (c) The same as panel (a), but for glasses formed by an infinitely fast quench, $\dot{T}\!\to\!\infty$. Note that the 3.6:1 and 3.2:1 scaling triangles correspond to the lines going through the $N\!=\!1600$ and $N\!=\!36$ data, respectively, demonstrating again a systematic variation of the exponent $\beta$ with $N$. (d) The same as panel (b), but for the data shown in panel (c).}
  \label{fig:fig1}
\end{figure*}

We simulate a generic glass-forming model in 2D --- also studied in~\cite{grzegorz_2d_modes_prl_2021,grzegorz_erratum_2022} --- in which point-like particles interact via an inverse-power-law potential that is smoothed at a cutoff distance up to two derivatives, see e.g.~\cite{cge_paper} for the model's details. We prepare ensembles of glasses by $(i)$ instantaneously quenching high-temperature liquid states to zero temperature (as done in~\cite{grzegorz_2d_modes_prl_2021,grzegorz_erratum_2022}), and $(ii)$ by annealing liquids for $10^4$ simulational time units (described, e.g., in~\cite{cge_paper}) at temperature $T\!=\!0.5$, which is approximately the (computer) glass transition temperature of our studied glass-forming model. Each of these annealing runs are followed by an instantaneous quench to zero temperature, to form a glass (termed ``well-annealed'' in Fig.~\ref{fig:fig1}). The ensemble sizes can be found in~\cite{footnote}.

\begin{figure*}[ht!]
  \includegraphics[width = 0.85\textwidth]{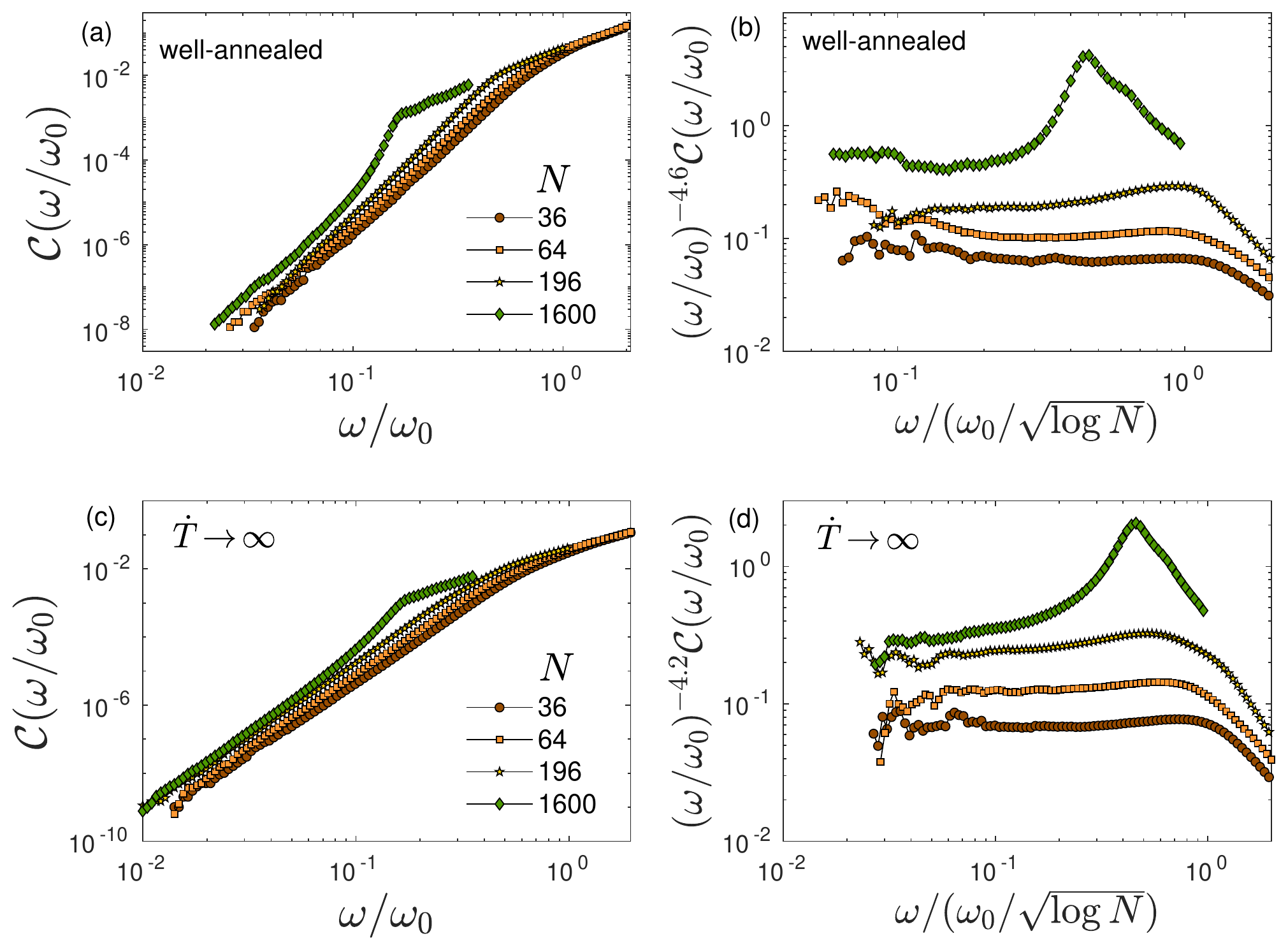}
  \caption{\footnotesize (a) The cumulative distributions ${\cal C}(\omega/\omega_0)\!\equiv\!\int_{0^+}^{\omega/\omega_0}{\cal D}(\omega'/\omega_0)d(\omega'/\omega_0)$ of well-annealed computer glasses plotted vs.~$\omega/\omega_0$, for various system sizes as indicated by the legend. (b) The same data as in panel (a), but here plotting $(\omega/\omega_0)^{-4.6}{\cal C}(\omega/\omega_0)$ vs.~$\omega/(\omega_0/\!\sqrt{\log{N}})$. (c) The same as panel (a), but for glasses formed by an infinitely fast quench, $\dot{T}\!\to\!\infty$ (d) Here we plot $(\omega/\omega_0)^{-4.2}{\cal C}(\omega/\omega_0)$ vs.~$\omega/(\omega_0/\!\sqrt{\log{N}})$ for the data shown in panel (c). A clear $N$-dependence of the exponent $\beta$ is seen in this representation.}
  \label{fig:fig2}
\end{figure*}

In Fig.~\ref{fig:fig1}, we present our results; panels (a) and (c) show the raw vibrational spectra ${\cal D}(\omega;N)$ of our pair of glass ensembles, respectively, as indicated in the figure legends. A clear system-size dependence is apparent in both ensembles. We further find that the exponent featured by the low-frequency power-law regime of ${\cal D}(\omega;N)$ drops below 4 for smaller, quickly quenched glasses, cf.~Fig.~\ref{fig:fig1}c, consistent with observations in 3D~\cite{lerner2019finite,protocol_prerc}. That is, we find that $\beta(N)\!<\!4$ increases with the system size $N$, from $\beta\!\simeq\!3.2$ for $N\=36$ to $\beta\!\simeq\!3.6$ for $N\=1600$.
In panels (b) and (d), we show that rescaling both axes according to the $N$ dependence of the nonphononic-excitations' characteristic scale $\sim\!1/\sqrt{\log N}$ as put forward in~\cite{modes_prl_2018,cge_paper} leads to a convincing data collapse, confirming that the nonphononic VDoS of 2D glasses is indeed $N$ dependent. In Fig.~\ref{fig:fig2}, we present the cumulative distributions ${\cal C}(\omega)\!\equiv\!\int_{0^+}^\omega{\cal D}(\omega')d\omega'$ showing again that $\beta$ features both glass-formation protocol and system-size dependencies. 
We note that the $N$ dependence of the exponent $\beta(N)$ also appears to be consistent with the results presented in the Supplemental Material file of~\cite{grzegorz_2d_modes_prl_2021,grzegorz_erratum_2022} (which can be found at~\cite{footnote2}). For example, in Fig.~2b therein the cumulative density of states normalized by $\omega^{4.5}$ is presented for various system sizes $N$, for a computer glass model closely related to the one used in this note. For $N\=3600$, the data feature a low-frequency plateau, which corresponds to $\beta\=4.5-1\=3.5$. For $N\!\le\!786$, however, the data curve up with decreasing $\omega$, indicating an exponent $\beta$ of value smaller than 3.5, consistently with the $N$ dependence of $\beta(N)$ we found.    

Our data demonstrate the difficulty to accurately determine the value of the exponent $\beta$ appearing in the low-frequency $\sim\!\omega^\beta$ scaling of the nonphononic spectrum of two-dimensional structural glasses. The proper determination of $\beta$ in the thermodynamic limit requires careful finite-size tests (see, e.g.~\cite{lerner2019finite} for three-dimensional glasses), which are particularly hard to conduct in two-dimensions due to the abundance of low-frequency phononic modes that obscure the nonphononic spectrum (see elaborate discussion on this issue in~\cite{phonon_widths}). In this context, we note that in~\cite{JCP_Perspective} systems of $N\!=\!25600$ particles were employed --- together with a nonlinear-excitation analyses~\cite{JCP_Perspective}---, in order to show (via simulations and extreme-value-statistics arguments) that $\beta\!=\!4$ in two-dimensional glasses, even if the latter were formed by an instantaneous quench from high-temperature liquid states. Our results, taken together with the data presented in~\cite{lte_pnas,modes_prl_2018,cge_paper,JCP_Perspective}, support that $\beta\!=\!4$ is the universal value characterizing the nonphononic spectrum of structural glasses in the thermodynamic limit, including in two-dimensions.

\newpage


\begin{thebibliography}{15}
\bibitem [{foo({\natexlab{a}})}]{footnote}%
  \BibitemOpen
  \bibinfo {note} {We created about 3.8M, 2.1M, 427K, and 94K independent,
  well-annealed glassy samples, and about 14M, 12M, 6.8M and 2.8M independent,
  instantaneously quenched glassy samples of sizes $N\!=\!36,64,196$, and 1600,
  respectively.}\BibitemShut {Stop}%
\end{thebibliography}
\end{document}